\documentstyle[12pt,tighten,aps,eqsecnum]{revtex}      
\begin{document}
\draft
\title{Simple method for calculating the Casimir energy \\ for sphere
}
\author{ V.~V.~Nesterenko\thanks{Electronic address:
nestr@thsun1.jinr.dubna.su}}
\address{Bogoliubov Laboratory of Theoretical Physics,
Joint Institute for Nuclear Research \\
Dubna, 141980, Russia}
\author{I.G.~Pirozhenko\thanks{Electronic
address: pirozhen@thsun1.jinr.dubna.su}}
\address{Petrozavodsk State University, Petrozavodsk,
185640, Russia}
\maketitle
\begin{abstract}

A simple method for calculating the Casimir energy for a sphere  is
developed which is based on a direct mode  summation and counter
integration in a complex  plane of eigenfrequencies. The method uses
only classical equations determining  the eigenfrequencies of the
quantum field
under consideration. Efficiency of this approach is demonstrated by
calculation of the Casimir energy for a perfectly conducting spherical
shell and fora massless scalar field obeying the Dirichlet and Neumann
boundary conditions on sphere. The possibility of rationalizing the
removal of divergences in this problem as a renormalization of
both the energy  and the radius of the sphere is discussed.
\end{abstract}
\section{Introduction}
\label{sec:Intr}
The Casimir effect attracts the attention of investigators during last
half of a century. More generally the Casimir effect
can be defined as an influence of the boundness of the
configuration space on the physical characteristics of the  quantum
field system (its energy, forces and momentum of forces acting on
the boundaries and so on).

This problem arises in different areas of theoretical physics:
in quantum electrodynamics (attractive force between two uncharged
conducting plates calculated by Casimir in 1948), in the theory of
elementary particles (the bag models of hadrons treat the energy
of quark and gluon fields located inside hadrons), in current
cosmology, in physics of condensed matter (elucidation of the
physical origin of sonoluminescence) and so on.

When considering the Casimir effect different methods are used:
Green function formalism~\cite{Milton}, stress-tensor
method~\cite{Deutsch},
multiple scattering
expansion~\cite{Balian}, zeta regularization
technique~\cite{Leseduarte},
heat-kernel series~\cite{Bordag}, direct mode summation with counter
integration~\cite{Lambiase,Nestr}.
Physical interpretations proposed in different approaches to
calculating the
Casimir effect are distinct. For example, the Casimir forces  between
two uncharged conducting plates can be treated both as the macroscopic
manifestation of the van der Waals forces and as an effect of zero
point oscillations of vacuum electromagnetic field. In this situation
it is worthwhile to separate in particular calculations an invariant,
with respect to physical interpretation, ``kernel'' which gives the
final
result.

In all the approaches to calculation of the Casimir effect a vague
point is the procedure of unique separation and subsequent removal
of the divergences. The lack of universal mathematically rigorous
prescription for this purpose leads in some problems to different
results
when different methods are applied (for example, calculation of the
Casimir energy for scalar massless field defined on the plane and
subjected to the Dirichlet boundary conditions on a
circle~\cite{Sen}).

With regard to all this,  the most simple, from mathematical point
of view, methods of calculation of the Casimir effect have an obvious
advantage because they right away allow one to reveal the
difficulties
generated by divergences. One of such methods is the direct summation
of eigenfrequencies of quantum field system by making use of counter
integration in complex frequency plane.
For the first time this approach was proposed
as a simple, in comparison with quantum field theory formalism,
method
of calculation  of the van der Waals forces between dielectrics
~\cite{Kampen}.  Further it was widely used in other
problems~\cite{Lambiase,Nestr,Br,Brev}.

The main goal of this paper is to show the simplicity and efficiency
of the direct mode summation by counter integration  when calculating
the Casimir energy for  such a difficult boundary as sphere.
This approach
is completely based on using the classical frequencies of quantum
field  system concerned, and the main tool employed is the Cauchy
theorem from
complex analysis. In this approach the Casimir energy for perfectly
conducting and infinitely thin spherical shell will be calculated.
Then the Casimir energy of scalar massless field subjected to the
Dirichlet or Neumann boundary conditions on sphere will be also
derived. As far as we know the last problem
(the Neumann boundary conditions)
has not been considered in other approaches yet. Unlike other authors
we propose to interpret the removal of divergences when calculating
the Casimir effect for sphere as the renormalization  not only of the
energy but also of the radius of the sphere.

The organization of the paper is as follows. In Sec.~\ref{sec:Spher}
we consider the vacuum energy of the electromagnetic field inside and
outside the perfectly conducting spherical shell.
In Sec.~\ref{sec:Dir} the Casimir  energy of massless scalar
field obeying the
Dirichlet or Neumann boundary conditions  on sphere is calculated.
Concluding
remarks and discussion of necessity for renormalization of the sphere
radius, in addition to the energy renormalization, are presented
in Sec.~\ref{sec:Concl}.
\section{Vacuum electromagnetic field inside and outside the perfectly
conducting spherical shell}
\label{sec:Spher}

The starting point of our approach is the following definition of the
Casimir energy
\begin{equation}
E=\frac{1}{2}\sum_{s}^{}\left(\omega_s-
\bar{\omega}_s\right).
\label{b1}
\end{equation}
Here $\omega_s$ are the eigenfrequencies of the system under
consideration,  and  $\bar{\omega}_s$ are those of the same
system, when  the parameters determining its boundaries take
on some limiting values. We need equations
for the oscillation frequencies of the electromagnetic field inside
and outside  the perfectly conducting sphere with radius $a$. There
are  two modes of oscillations: transverse-electric modes
and transverse magnetic  ones (TE-modes and TM-modes, respectively).
The eigenfrequencies of the TE-modes are defined by the
equations~\cite{Stratton}
\begin{eqnarray}
 j_l(\omega a)&=&0,\label{b2}\\
 h_l^{(1)}(\omega a)&=&0,\label{b3}
\end{eqnarray}
and the eigenfrequencies for the TM-modes are given by
\begin{eqnarray}
\frac{d}{dr}\left.\left[r j_l(\omega r) \right]\right|_{r=a}&=&0,
\label{b4}\\
\frac{d}{dr}\left.\left[r h_l^{(1)}(\omega r) \right]
\right|_{r=a}&=&0.
\label{b5}
\end{eqnarray}
In formulae (\ref{b2})-(\ref{b5}) $j_l(z)$ and $h_l^{(1)}(z)$ are
the spherical Bessel functions~\cite{Abram}
\begin{equation}
j_l(z)=\sqrt{\frac{\pi}{2 z}}J_{l+1/2}(z),\;\;\;
h_l^{(1)}(z)=\sqrt{\frac{\pi}{2 z}}H^{(1)}_{l+1/2}(z),
\label{b6}
\end{equation}
and $l=1,2,\dots$. Only positive roots of these equations
$\omega_{nl}>0,\;n=1,2,\dots$ should be considered.
Equations (\ref{b2}) and (\ref{b4}) specify the frequencies of the
electromagnetic  oscillations inside the sphere and
Eqs. (\ref{b3}) and (\ref{b5}) give the frequencies outside
the sphere~\cite{Stratton}.

In the case of spherical boundary the sum $\sum_{s}^{}$  in (\ref{b1})
can be written as
\begin{equation}
\frac{1}{2}\sum_{s}^{}\omega_{s}=\frac{1}{2}\sum_{l=1}^{\infty}
\sum_{m=-l}^{l}\sum_{n=1}^{\infty}\omega_{nl}=\sum_{l=1}^{\infty}
(l+1/2) S_{l},
\label{b7}
\end{equation}
Where $S_{l}=\sum_{n=1}^{\infty}\omega_{nl}$, and each frequency
equation
(\ref{b2})--(\ref{b5}) generates its partial sum $S_{l}^{\alpha},\;
\alpha=1,\dots,4$.

For the partial sums $S_{l}^{(\alpha)}$ we use integral representation
that follows from the Cauchy theorem~\cite{Wittaker}
\begin{equation}
S_{l}^{(\alpha)}=\frac{1}{2\pi i}\oint_{C}dz\,z \frac{d}{dz}
\ln  f^{(\alpha)}(z,a).
\label{b8}
\end{equation}
Here $f^{(\alpha)}(z,a)$ are the functions defining the frequency
equations (\ref{b2})--(\ref{b5}) in the form
\begin{equation}
f^{(\alpha)}(\omega,a)=0,\;\;\alpha=1,2,3,4.
\label{b9}
\end{equation}
The counter $C$ encloses counterclockwise positive roots of these
equations.
Taking into account the position of the roots on real axis one can
deform the counter  $C$ in such a way that it will consist of the
segment
$[-i \Lambda,\;i\Lambda]$ of
the imaginary axis and a semicircle of radius~$\Lambda$
with $\Lambda \to \infty$
in the right half-plane.
When $\Lambda$ is fixed, the counter integral (\ref{b8}) gives
the regularized value of corresponding frequency sum (it sums up
the finite number of the roots of the frequency equation (\ref{b8})
that lie inside the counter).

From physical considerations it is clear that for negative values
of the argument $\omega$ the functions $f^{(\alpha)}(\omega,a)$
have to be defined by a condition
\begin{equation}
f^{(\alpha)}(-\omega,a)=f^{(\alpha)}(\omega,a),\;\;\;\omega>0.
\label{b10}
\end{equation}
This can be achieved, for example, by introducing a finite photon mass
which is equated to zero at the end of the calculations.
Separating the contributions of different parts of the counter $C$,
we can rewrite formula (\ref{b8}) as
\begin{eqnarray}
S_{l}^{(\alpha)}&=&-\frac{1}{2\pi}\int\limits_{-\Lambda}^{+\Lambda}
dy\,y\,
\frac{d}{dy}\ln  f^{(\alpha)}(i y,a)\nonumber\\
&&+\frac{1}{2\pi i}
\int\limits_{C_{\Lambda}}^{}z\,d\ln  f^{(\alpha)}(z,\alpha).
\label{b11}
\end{eqnarray}
Here $C_{\Lambda}$ is the semicircle of radius  $\Lambda$ introduced
above. In the first term on the right-hand side
of (\ref{b11}) we can integrate by parts, the nonintegral term
being omitted in view of (\ref{b10}).
On the other hand this term can be removed by a subtraction which
we shall discuss further.

In accordance with the definition (\ref{b1})
in order to obtain a finite (observable) value of the Casimir
energy it is necessary to perform the subtraction. As usual, we shall
subtract the contribution of the Minkowski space that corresponds
to the limit $a=\infty$ in Eq.~(\ref{b11}). Letting
$\bar{S}_{l}^{(\alpha)}$ represent the value of the partial
sum $S_{l}^{(\alpha)}$ which is to be subtracted from
(\ref{b11}) we get
\begin{eqnarray}
\bar{S}_{l}^{(\alpha)}&=&\frac{1}{2\pi}\int
\limits_{-\Lambda}^{+\Lambda}
dy\,\ln  f^{(\alpha)}(iy,a\to\infty)\nonumber\\
&&+\frac{1}{2\pi i} \int\limits_{C'_{\Lambda}}^{}
z\,d \ln  f^{(\alpha)}(z,a\to\infty).
\label{b12}
\end{eqnarray}
In view of an  oscillating character of the
function $f^{(\alpha)}(z,a)$  (see below)
we have on the semicircle $C_{\Lambda}$
\begin{equation}
\lim_{a\to\infty}f^{(\alpha)}(z,a)=f^{(\alpha)}(z,a)
\label{b13}
\end{equation}
Thus, for the difference $S_l^{(\alpha)}-\bar{S}_l^{(\alpha)}$
in the definition of the Casimir energy (\ref{b1}), we obtain
\begin{equation}
S_l^{(\alpha)}-\bar{S}_l^{(\alpha)}=\frac{1}{\pi}
\int\limits_{0}^{\infty}
dy\,\ln \left[
\frac{f^{(\alpha)}(iy,a)}{f^{(\alpha)}(iy,a\to\infty)}\right].
\label{b14}
\end{equation}
Here again the property (\ref{b10}) has been used.
Now we proceed to substituting into Eq.~(\ref{b14}) the concrete
expressions for the functions $f^{(\alpha)}$ defined by
frequency
equations (\ref{b2})--(\ref{b5}).
From Eq.~(\ref{b2}) we have
\begin{eqnarray}
\frac{f^{(1)}(iy,a)}{f^{(1)}(iy,a\to\infty)}&=&
\frac{J_{\nu}(iya)}{ \lim\limits_{a\to\infty}J_{\nu}(iya)}=
\frac{I_{\nu}(ay)}{ \lim\limits_{a\to\infty}I_{\nu}(ay)}\nonumber\\
&=&\sqrt{2\pi a y}\, e^{-ay}I_{\nu}(ay),
\label{b15}
\end{eqnarray}
where $\nu=l+1/2$, and
$I_{\nu}(z)$ is the modified Bessel function
$J_{\nu}(i z)=i^{\nu}I_{\nu}(z)$. We have used here the asymptotics
of the
function $I_{\nu}(z)$ for fixed value of  $\nu$ and large
$z$\cite{Abram}
\begin{equation}
I_{\nu}(z)\simeq\frac{e^z}{\sqrt{2\pi z}}.
\label{b16}
\end{equation}

From frequency equation  (\ref{b3}) it follows that
\begin{equation}
\frac{f^{(2)}(iy,a)}{f^{(2)}(iy,a\to\infty)}=
\frac{H_{\nu}^{(1)}(iay)}{\lim\limits_{a\to\infty}H_{\nu}^{(1)}(iay)},
\label{b17}
\end{equation}
where $H_{\nu}^{(1)}=J_{\nu}(z)+i N_{\nu}(z)$ is the Hankel function
of the first kind. Using the modified Bessel functions
$ K_{\nu}(z)=(\pi/2)i^{\nu+1} H_{\nu}^{(1)}(i z)$ we rewrite
Eq.~(\ref{b17})
as
\begin{equation}
\frac{f^{(2)}(iy,a)}{f^{(2)}(iy,a\to\infty)}=
\frac{K_{\nu}(ay)}{\lim\limits_{a\to\infty}K_{\nu}(ay)}
=\sqrt{\frac{2 a y}{\pi}}e^{a y}K_{\nu}(ay).
\label{b18}
\end{equation}
Here we employed the asymptotics of the function $K_{\nu}(z)$ for
large $z$ and fixed $\nu$~\cite{Abram}
\begin{equation}
K_{\nu}(z)\simeq\sqrt{\frac{\pi}{2 z}}e^{-z}.
\label{b19}
\end{equation}

Thus, the TE-modes give the following contribution to Eq.~(\ref{b1})
\begin{equation}
\sum_{\alpha=1}^{2}(S_l^{(\alpha)}-\bar{S}_l^{(\alpha)})=
\frac{1}{\pi}\int\limits_{0}^{\infty}dy\,
\ln \left[2ay I_{\nu}(ay) K_{\nu}(ay)\right].
\label{b20}
\end{equation}

In the same way we deduce from the frequency equation (\ref{b4})
\begin{eqnarray}
\frac{f^{(3)}(iy,a)}{f^{(3)}(iy,a\to\infty)}&=&
\frac{J_{\nu}(iya)/2+i y a J_{\nu}'(i y a)}{
\lim\limits_{a\to\infty}[J_{\nu}(iya)/2 +i y a J_{\nu}'(i y
a)]}\nonumber\\
&=&\frac{I_{\nu}(ya)/2+ y a I_{\nu}'(y a)}{
\lim\limits_{a\to\infty}[ I_{\nu}(ya)/2 + y a I_{\nu}'( y
a)]}
\label{b21}
\end{eqnarray}
The prime over the Bessel function means the differentiation with
respect  to its argument.
From (\ref{b16}) it follows that
\begin{equation}
\lim_{a\to\infty}\left[I_{\nu}(a y)/2+a y I_{\nu}'(a y)\right]
=\sqrt{\frac{a y}{2 \pi}} e^{a y}.
\label{b22}
\end{equation}
In view of this, formula (\ref{b21}) assumes the form
\begin{equation}
\frac{f^{(3)}(iy,a)}{f^{(3)}(iy,a\to\infty)}=
\sqrt{\frac{2 \pi}{a y}}e^{-a y}\left[I_{\nu}(a y)/2+
a y\,I_{\nu}'(ay)\right].
\label{b23}
\end{equation}

In  the same way, for the frequency equation (\ref{b5}) we obtain
\begin{eqnarray}
\lefteqn {\frac{f^{(4)}(i y,a)}{f^{(4)}(i y,a\to\infty)}} \nonumber \\
&=&\frac{H^{(1)}_{\nu}(i y a)/2+
ia y {H_{\nu}^{(1)}}'(i a y)}{\lim\limits_{a\to\infty}
\left[ H_{\nu}^{(1)}(iay)/2+i a y {H_{\nu}^{(1)}}'(iay)\right]}
\nonumber \\
&=&\frac{K_{\nu}(a y)/2+a y K'_{\nu}(a y)}{\lim
\limits_{a\to\infty}\left[K_{\nu}(a y)/2+a y K'_{\nu}(a y)\right]}
\nonumber \\
&=&-\sqrt{\frac{2}{\pi y a}}e^{ay}\left[K_{\nu}(ay)/2+
a y K'_{\nu}(a y)\right].
\label{b24}
\end{eqnarray}
Summing up Eq.~(\ref{b23}) and (\ref{b24}), we arrive at the
contribution of the  TM-modes
\begin{eqnarray}
\lefteqn{\sum_{\alpha=3}^{4}(S_{l}^{(\alpha)}-\bar{S}_{l}^{(\alpha)})}
\nonumber\\
&=&\frac{1}{\pi}\int\limits_{0}^{\infty}dy\,\ln \left\{-\frac{2}{ay}
\left[\frac{1}{2}I_{\nu}(ay)+a y {I}_{\nu}'(ay)\right]\right.
\nonumber\\
&&\left.\times\left[\frac{1}{2}K_{\nu}(ay)+a y
{K}_{\nu}'(ay)\right]\right\}.
\label{b25}
\end{eqnarray}

Finally for the Casimir energy (\ref{b1}) we obtain from (\ref{b7}),
(\ref{b20}) and (\ref{b25})
\begin{equation}
E=\frac{1}{\pi a}\sum_{l=1}^{\infty}\left(l+\frac{1}{2}\right)
\int\limits_{0}^{\infty}dy
\,\ln \left[1-\left({\sigma}'_l(y)\right)^2\right],
\label{b26}
\end{equation}
where the notation
\begin{eqnarray*}
1-\left({\sigma}'_l(y)\right)^2&=&-4 I_{\nu}(y)K_{\nu}(y)\\
&&\times\left[\frac{1}{2}I_{\nu}(y)+y {I}'_{\nu}(y)\right]
\left[\frac{1}{2}K_{\nu}(y)+y {K}'_{\nu}(y)\right]
\end{eqnarray*}
is introduced. Using the value of the Wronskian of
the modified Bessel functions
$I_\nu (y)$ and $K_\nu (y)$~\cite{Abram}
$$
I_\nu (y) K'_\nu (y)- I_\nu '(y)K_\nu (y) = -\frac{1}{y}
$$
one can show that
$$ \sigma_{l}(y)=y I_{\nu}(y) K_{\nu}(y),\quad \nu=l+1/2.
$$
The integral in (\ref{b26}) converges. This
follows from the asymptotics of $\sigma'_l(y)$
for large $y$ and fixed $\nu=l+1/2$~\cite{Abram}
\begin{equation}
{\sigma}'_l(y)\simeq-\frac{1}{2 y^2}\left[1-
\frac{1}{2}\frac{4\nu^2-1}{(2 y)^2}+\dots\right].
\label{b27}
\end{equation}
Formula  (\ref{b26})  coincides with Eq.~(5.1) in paper~\cite{Milton},
on the condition that the cut-off factor in the last equation
is omitted and the
integration by parts is performed.
To carry out the summation with respect to $l$ in (\ref{b26}) one
needs the
 behavior of the integral
\begin{equation}
Q_l=\frac{l+1/2}{\pi}\int\limits_{0}^{\infty}d y \,
\ln \left[1-\left({\sigma}'_l(y)\right)^2\right]
\label{b28}
\end{equation}
at large $l$.
Applying the uniform with respect to $z$ asymptotics for the modified
Bessel functions at large $\nu$~\cite{Milton,Abram}
\begin{equation}
I_{\nu}(\nu z)K_{\nu}(\nu z)\simeq\frac{1}{2\nu}
\frac{1}{(1+z^2)^{1/2}},
\label{b29}
\end{equation}
we obtain from (\ref{b28})
\begin{eqnarray}
Q_l&\simeq&\frac{\nu^2}{\pi}\int\limits_{0}^{\infty}
dz\,\ln \left[1-\frac{1}{4 \nu^2(1-z^2)^3}\right]\nonumber\\
&\simeq&-\frac{1}{4\pi}\int\limits_{0}^{\infty}\frac{dz}{(1+z^2)^3}
=-\frac{3}{64}, \quad l \to \infty.
\label{b30}
\end{eqnarray}
Thus, the sum (\ref{b26}) at large $l$ diverges as
$\sum_{l=1}^{\infty}(l+1/2)^0$.
To determine the finite value for this sum  we rewrite (\ref{b26})
in the following way
\begin{eqnarray}
E&=&\frac{1}{a}\sum_{l=1}^{\infty}\left[Q_l+\frac{3}{64}
-\frac{3}{64}\right]\nonumber\\
&=&\frac{1}{a}\sum_{l=1}^{\infty}\bar{Q}_l-\frac{3}{64 a}
\sum_{l=1}^{\infty}\left(l+\frac{1}{2}\right)^0,
\label{b31}
\end{eqnarray}
where
\begin{equation}
\bar{Q}_l=Q_l+3/64.
\label{b32}
\end{equation}
The sum $\sum_{l=1}^{\infty}\bar{Q}_l$  converges
because at large $l$
\begin{equation}
\bar{Q}_l=-\frac{9}{16384 \nu^2}+{\cal O}(\nu^{-4}).
\label{b33}
\end{equation}
The last divergent sum
in (\ref{b31}) can be defined by using the Hurwitz zeta
function~\cite{Wittaker}
\begin{equation}
\zeta(z,q)=\sum_{n=0}^{\infty}\frac{1}{(q+n)^z},
\label{b34}
\end{equation}
which at  $q=1/2$ is related with the Riemann
$\zeta$-function~\cite{Rizhik}
\begin{equation}
\zeta(z,1/2)=(2^z-1)\zeta(z).
\label{b35}
\end{equation}
From (\ref{b34}) it follows that
\begin{equation}
-\frac{3}{64 a}\sum_{l=1}^{\infty}\left(l+\frac{1}{2}\right)^0=
-\frac{3}{64 a}\left(\zeta(0,1/2)-1\right)=\frac{3}{64 a}
\label{b36}
\end{equation}
since $\zeta(0,1/2)=0$.

Finally for the Casimir energy we obtain
\begin{equation}
E=\frac{1}{a}\sum_{l=1}^{\infty}\bar{Q}_l+\frac{3}{64 a},
\label{b37}
\end{equation}
where $\bar{Q}_l$ is defined in (\ref{b32}) and (\ref{b28}).
The sum $\sum_{l}^{}\bar Q_l$ in the right-hand side of (\ref{b37})
can be estimated with allowance for the asymptotics~(\ref{b33})
\begin{eqnarray}
\sum_{l=1}^{\infty}\bar Q_l&\simeq&-\frac{9}{16384}
\sum_{l=1}^{\infty}\frac{1}{(l+1/2)^2}=-\frac{9}{2^{14}}
[\zeta(2,1/2)-4]
\nonumber\\
&=&-\frac{9}{2^{14}}\left(\frac{\pi^2}{2}-4\right)=-0.000514\dots.
\label{b38}
\end{eqnarray}
Thus, the main contribution to (\ref{b37}) is given by the second
term
and to a good approximation one can put for the Casimir energy
\begin{equation}
E\simeq\frac{3}{64a}=\frac{1}{a}\,0.046875.
\label{b39}
\end{equation}
Taking into account (\ref{b38}) we get instead of (\ref{b39})
$$E\simeq\frac{1}{a}\,0.046361\dots.$$
With greater accuracy this energy has been calculated
in~\cite{Milton}.

\section{Scalar field obeying the Dirichlet or Neumann boundary
conditions on sphere}
\label{sec:Dir}

The suggested method can be easily applied to the calculation of the
Casimir
energy of a massless scalar field subjected to boundary conditions
on a
sphere. Let us first consider  the Dirichlet boundary conditions.
In this case the eigenfrequencies are given by Eqs.~(\ref{b2}) and
(\ref{b3})
with $l=0,1,2\dots$.
From (\ref{b1}), (\ref{b7}) and (\ref{b20}) we obtain
\begin{eqnarray}
E^{({\cal D})}&=&\sum_{l=0}^{\infty}\left(l+\frac{1}{2}\right)
\frac{1}{\pi}
\int\limits_{0}^{\infty}dy\,\ln [2 a y I_{\nu}(ay)K_{\nu}(ay)]
\nonumber\\
&=&\frac{1}{a}\sum_{l=0}^{\infty}\left(l+\frac{1}{2}\right)
\frac{1}{\pi}
\int\limits_{0}^{\infty}dy \,\ln [2 y I_{\nu}(y) K_{\nu}(y)],
\label{c1}\\
\nu&=&l+1/2.\nonumber
\end{eqnarray}

It is worth comparing this expression with Eq.~(3.5), derived
in paper~\cite{Bender} by the Green function method. The essential
advantage of our approaches is  absence in the integrand of the
constant terms
which lead to the divergences (so-called contact terms).
For large $\nu$ the integral in (\ref{c1}) behaves as follows
\begin{eqnarray}
Q_l&\equiv&\frac{l+1/2}{\pi}\int\limits_{0}^{\infty}dy\,
\ln [2 y I_{\nu}(y)K_{\nu}(y)]\nonumber\\
&\simeq&-\frac{{\nu}^2}{2}-\frac{1}{128}+\frac{35}{32768{\nu}^2}+
{\cal O}({\nu}^{-3}).
\label{c2}
\end{eqnarray}
The first two terms in (\ref{c2}) give rise to divergences when
summing up
with respect to
$l$ in (\ref{c1}). These divergences can be removed as in the
previous section  by means of the Hurwitz
$\zeta$-function  (\ref{b34}). On this purpose
we rewrite~(\ref{c1})
\begin{eqnarray}
E^{({\cal D})}&=&\frac{1}{a}\sum_{l=0}^{\infty}Q_l\nonumber\\
&=&\frac{1}{a}\sum_{l=0}^{\infty}\left(Q_l+\frac{\nu^2}{2}+
\frac{1}{128}\right)\nonumber\\
&&-\frac{1}{2 a}\sum_{l=0}^{\infty}\left(l+
\frac{1}{2}\right)^2-\frac{1}{128 a}\sum_{l=0}^{\infty}\left(l+
\frac{1}{2}\right)^0.
\label{c3}
\end{eqnarray}
Here we have added and subtracted under the sum sign two first terms
of asymptotic expansion (\ref{c2}). Taking into account
(\ref{b34}) we obtain
\begin{eqnarray}
E^{({\cal D})}&=&\frac{1}{a}\sum_{l=0}^{\infty}\bar{Q}_l-\frac{1}{2 a}
\zeta\left(-2,\frac{1}{2}\right)-
\frac{1}{128 a}\zeta\left(0,\frac{1}{2}\right)\nonumber\\
&=&\frac{1}{a}\sum_{l=0}^{\infty}\bar{Q}_l,
\label{c4}
\end{eqnarray}
where
\begin{equation}
\bar{Q}_l=Q_l+\frac{1}{2}\left(l+\frac{1}{2}\right)^2+\frac{1}{128}.
\label{c5}
\end{equation}
We derived the last equality in (\ref{c4}) bearing in mind
that $\zeta(-2,1/2)=0$ and $\zeta(0,1/2)=0$.
By virtue of the asymptotics (\ref{c2}) the last sum in (\ref{c4})
obviously converges.

With increase of $\nu$, the modified Bessel functions $I_{\nu}(z)$
and $K_{\nu}(z)$
are rapidly approaching their uniform asymptotics.
That is why even for comparatively small values of $l$ we can assume
with allowance for (\ref{c2}) and (\ref{c5})
\begin{equation}
\bar{Q}_l\simeq\bar{Q}_l^{asym}=\frac{35}{32768{\nu}^2}.
\label{c6}
\end{equation}
This simplifies the numerical calculations considerably.
It is important to note that direct calculation of the
modified  Bessel functions $I_{\nu}(z)$ and
$K_{\nu}(z)$ with desired accuracy for all $z$ and at large $\nu$
is a technically difficult problem~\cite{Brevik}. It is especially
concerns the product $I_{\nu}(z)K_{\nu}(z)$ encountered in the
integrand of (\ref{c2}).

Table \ref{table1} demonstrates  the applicability of the formula
(\ref{c6}).
Calculating $\bar{Q}_l$ for $l\le 3$ by means of
numerical integration of the expressions including the product
$I_{\nu}(y)K_{\nu}(y)$ (see Eqs.~(\ref{c5}) and (\ref{c2})) and
using the
asymptotic formula (\ref{c6}) for $l>3$ we obtain
\begin{equation}
E^{(\cal D)}=\frac{1}{a}\,0.002819\dots.
\label{c7}
\end{equation}
With greater accuracy the Casimir energy for spherical conducting
shell
was calculated in~\cite{Bender} by making use of the Green function
technique.

Now we proceed to the consideration of the energy of a scalar field
obeying the Neumann boundary conditions on the sphere. The
eigenfrequencies
inside and outside the sphere are defined now by equations
\begin{eqnarray}
\frac{d}{dr}\left[j_l(\omega r)\right]\left|_{r=a}\right.&=&0,
\label{c8}\\
\frac{d}{dr}\left.\left[h^{(1)}_l(\omega r)\right]\right|_{r=a}&=&0,
\label{c9}
\end{eqnarray}
where $l$ takes values $0,1,2,\dots$. By analogy with
Eq. (\ref{b25}) we derive
\begin{eqnarray}
\lefteqn{E^{(\cal N)}=}\nonumber\\
&=&\sum_{l=0}^{\infty}\left(l+\frac{1}{2}\right)
\frac{1}{\pi}\int\limits_{0}^{\infty}dy\,\ln \left\{-\frac{2}{ay}
\right.
\left[\frac{1}{2}I_{\nu}(ay)-a y {I}'_{\nu}(ay)\right]
\nonumber\\
&&\times\left.
\left[\frac{1}{2}K_{\nu}(ay)-a y {K}'_{\nu}(ay)\right]\right\}.
\label{c10}
\end{eqnarray}
As in the case of electromagnetic field it is convenient at first
to consider the sum
$E^{(\cal D)}+E^{(\cal N)}$. From (\ref{c1}) and (\ref{c10}) it
follows
that
\begin{eqnarray}
\lefteqn{E^{(\cal D)}+E^{(\cal N)}}\nonumber\\
&=&\frac{1}{\pi a}\sum_{l=0}^{\infty}
\left(l+\frac{1}{2}\right)\int\limits_{0}^{\infty}dy\,
\ln \left[1-\left(\mu_l(y)\right)^2\right],
\label{c11}
\end{eqnarray}
where the notation
\begin{eqnarray}
1-(\mu_l(y))^2&\equiv&-4 I_{\nu}(y)K_{\nu}(y)
\left[\frac{1}{2}I_{\nu}(y)-y {I}'_{\nu}(y)\right]\nonumber\\
&&\times\left[\frac{1}{2}K_{\nu}(y)-y {K}'_{\nu}(y)\right]
\label{c12}
\end{eqnarray}
is introduced.
It is easy to show that
\begin{equation}
\mu_l(y)=y^2\frac{d}{dy}\left(\frac{1}{y}I_{\nu}(y)K_{\nu}(y)\right).
\label{c13}
\end{equation}
The integral in (\ref{c11}) converges since for large $y$ and fixed
$l$ the following estimation
\begin{equation}
\mu_l(y)\simeq-\frac{1}{y},\;\;y\to\infty
\label{c14}
\end{equation}
holds.

The convergence of the sum in (\ref{c11}) is determined by the
behavior at large $l$ of the expression
\begin{equation}
P_l=\frac{l+1/2}{\pi}\int\limits_{0}^{\infty}dy \ln \left[1-
(\mu(y))^2\right].
\label{c15}
\end{equation}
Using the uniform asymptotics of the modified Bessel
functions~\cite{Abram},
we obtain
\begin{equation}
P_l\simeq-\frac{19}{64}-\frac{153}{16384\nu^2}+{\cal O}(\nu^{-3}).
\label{c16}
\end{equation}

The first term in this expression leads to a divergence in (\ref{c11})
when summing with respect to $l$. We again overcome this difficulty
using the  Hurwitz zeta function
\begin{eqnarray}
E^{(\cal D)}+E^{(\cal N)}&\equiv&\frac{1}{a}\sum_{l=0}^{\infty}P_l
=\frac{1}{a}\sum_{l=0}^{\infty}\bar{P}_l-\frac{19}{64 a}
\zeta\left(0,\frac{1}{2}\right)\nonumber\\
&=&\frac{1}{a}\sum_{l=0}^{\infty}
\bar{P}_l,
\label{c17}
\end{eqnarray}
where $\bar{P}_l$ is the renormalized value of $P_l$
\begin{equation}
\bar{P}_l=P_l+\frac{19}{64}
\label{c18}
\end{equation}

When calculating the last sum in (\ref{c17}) we can again assume
\begin{equation}
\bar{P}_l\simeq\bar{P}_l^{asym}=-\frac{153}{16384\nu^2},\;\;\nu=l+1/2.
\label{c19}
\end{equation}

Numerical calculation shows (see Table~\ref{table2}) that with
increasing $l$
$P_l$ approaches rapidly to its asymptotic value defined by
Eq.~(\ref{c19}). If the required accuracy is not too high one can
use Eq.~(\ref{c19}) even for $l\ge3$. This leads to the result
\begin{equation}
E^{(\cal D)}+E^{(\cal N)}=-\frac{1}{a}\,0.220958\dots\,{.}
\label{c20}
\end{equation}
Hence, taking into account (\ref{c7}) we have for the Casimir energy
of massless scalar field obeying the Neumann boundary conditions
on the sphere
\begin{equation}
E^{(\cal N)}=-\frac{1}{a}\,0.223777\dots
\label{c21}
\end{equation}
As far as we know this result is obtained here for the first time.
\section{Conclusion}
\label{sec:Concl}

The direct summation of eigenfrequencies by calculation of the Casimir
effect for nonflat boundaries (specifically for sphere) has been used
only in pioneer paper by Boyer~\cite{Boyer}.
The fact that done by us is actually a development and maximum
simplification of the Boyer method and bringing it  to such a form
when  numerical calculations are practically not required
(see Eq.~(\ref{b39})),
and, what is more important, cut-off functions are not used.
In other approaches, for example, by making use of the Green
function formalism~\cite{Milton,Bender}, transition to the
imaginary
frequencies is used without detailed justification.
Contour integration
in the mode summation method supplies a clear explanation
for this technical
trick.

Let us turn now to  removing the divergences in the problem under
consideration.
Once renormalization of the Casimir energy by the formula
(\ref{b1}) is accomplished the divergence for all that remains.
In the general case the formula concerned has the form
\begin{equation}
E=\frac{C_1+C_2^{\infty}}{a},
\label{d1}
\end{equation}
where $C_1$ is finite constant, and $C_2$ is a divergent expression.
Thus for example, for the electromagnetic field
$C_1=0.046176\dots$ and $C_2^{\infty}$ is given by the divergent
series
\begin{equation}
C_2^{\infty}=-\frac{3}{64}\sum_{l=1}^{\infty}(l+1/2)^0.
\label{d2}
\end{equation}
To remove this divergence we have applied the formal technique
of the zeta function renormalization.
In Ref.~\cite{Milton} the finite result was obtained here by making
use   of an  exponential cutting multiplier splitting the
arguments of the field
operators in energy-momentum tensor.  When calculating the Casimir
energy for scalar massless field obeying the Dirichlet boundary
conditions on sphere
$C_1=0.002819\dots$ and $C_2^{\infty}$  stands for the sum of the
divergent
series
\begin{equation}
C_2^{\infty}=-\frac{1}{2}\sum_{l=0}^{\infty}(l+1/2)^2
-\frac{1}{128}\sum_{l=0}^{\infty}(l+1/2)^0.
\label{d3}
\end{equation}
Both in our paper and in  Ref.~\cite{Bender} this divergence has been
taken away by formal technique of the Hurwitz $\zeta$-function.

Here the following   question arises: to renormalization of which
parameter does the removal of the divergence
$C_2^{\infty}$ correspond? After renormalization of the Casimir
energy according to Eq.~(\ref{b1}) only one parameter, namely, radius
of the  sphere  $a$, is available. Therefore it is natural to treat
the
removal of the divergence
$C_2^{\infty}$ as the renormalization of the sphere radius. This can
be done in a standard way by transition from the initial (bare)
radius $a$ to the physical (observable) radius
$a_{phys}$:  $a=a_{phys}+\delta a$, where $\delta a$
is the appropriate counterterm. In view of this, Eq.~(\ref{d1}) can
be rewritten  as
\begin{equation}
E=\frac{C_1+C_2^{\infty}}{a}=\frac{C_1+C_2^{\infty}}{a_{phys}+
\delta a}=\frac{C_1}{a_{phys}}\frac{\left(1
+C_2^{\infty}/C_1\right)}{\left(1+\delta a/a_{phys}\right)}.
\label{d4}
\end{equation}
Setting
\begin{equation}
\frac{\delta a}{a_{phys}}=\frac{C_2^{\infty}}{C_1},
\label{d5}
\end{equation}
one arrives at the finite result
\begin{equation}
E=\frac{C_1}{a_{phys}}.
\label{d6}
\end{equation}

This reasoning seems to be more consistent as compared with, for
example, introduction  into consideration of a phenomenological
interaction~\cite{Sen}
localized on sphere with subsequent renormalization of the  coupling
constant of this interaction that finally takes up  divergence
$C_2^{\infty}$. It stands no reason that explanation  suggested
is also applicable to the calculation of the Casimir effect for the
field
confined inside the cavity. Certainly, in this case the values of
the
constants  $C_1$ and
$C_2^{\infty}$ in Eq.~(\ref{d1}) will be different.

\acknowledgments

One of the authors (V.V.N.) considers it his duty to thank
Prof.\ K.~A.\ Milton for valuable discussions of many topics
concerned in the paper and Prof.\ I.~H.\ Brevik for interest in
the work.

This work was accomplished with financial support of Russian
Foundation
of Fundamental Research (grant ü 97-01-00745).

\begin{table}
\caption{ $\bar{Q}_l$ obtained by numerical integration according to
Eqs.~(3.2), (3.5), and asymptotic formula (3.6) for
$\bar{Q}_l^{asym}$. }
\label{table1}
\begin{tabular}{lcr}
$l$&$\bar{Q}_l$&$\bar{Q}_l^{asym}$\\
\tableline
0&0.001913&0.004273\\
1&0.000398&0.000474\\
2&0.000159&0.000171\\
3&0.000084&0.000087\\
4&0.000052&0.000053
\end{tabular}
\end{table}

\begin{table}
\caption{ With increasing $l$  $\bar{P}_l$ rapidly  approaches to its
asymptotic value $ \bar{P}_l^{asym} $ given in (3.19). }
\label{table2}
\begin{tabular}{lcr}
$l$&$\bar{P}_l$&$\bar{P}_l^{asym}$\\
\tableline
0&-0.211491&-0.037351\\
1&-0.004800&-0.004150\\
2&-0.001582&-0.001494\\
3&-0.000775&-0.000762
\end{tabular}
\end{table}

\end{document}